\documentclass[10pt]{article}
\usepackage[OE]{express}

\begin{document}

\title{Giant enhancement of reflectance due to the interplay between surface confined wave modes and nonlinear gain in dielectric media}

\author{Sangbum Kim and Kihong Kim\authormark{*}}

\address{Department of Energy Systems
Research and Department of Physics, Ajou University, Suwon 16499, Korea}

\email{\authormark{*}khkim@ajou.ac.kr}

\begin{abstract}
We study theoretically the interplay between the surface confined wave modes and
the linear and nonlinear gain of the dielectric layer in the Otto
configuration. The surface confined wave modes such as surface
plasmons or waveguide modes are excited in the dielectric-metal
bilayer by obliquely incident $p$ waves. In the purely linear case,
we find that the interplay between linear gain and surface confined
wave modes can generate a large reflectance peak with its value much
greater than 1. As the linear gain parameter increases, the peak
appears at smaller incident angles, and the associated modes also
change from surface plasmons to waveguide modes. When the nonlinear
gain is turned on, the reflectance shows very strong multistability
near the incident angles associated with surface confined wave
modes. As the nonlinear gain parameter is varied, the reflectance
curve undergoes complicated topological changes and sometimes
displays separated closed curves. When the nonlinear gain parameter
takes an optimally small value, a giant amplification of the
reflectance by three orders of magnitude occurs near the incident
angle associated with a waveguide mode. We also find that there exists a range of
the incident angle where the wave is dissipated rather than
amplified even in the presence of gain. We suggest that this can
provide the basis for a possible new technology for thermal control
in the subwavelength scale.
\end{abstract}


\section{Introduction}

Nanophotonics is the study of the interaction of light
and matter on the subwavelength scale and is a rapidly growing
field in nanoscience. Surface plasmon polaritons consist of
the light waves trapped on the surface of a medium and the resonant
oscillations of surface charges with the light waves. They lead to
the enhancement of electromagnetic fields near the surface, thus are
suitable for applications in sensor technology
\cite{Homola-etal-SAB1999,Kneipp-etal-JPC2002}. The applications
include probing thin films \cite{Pockrand-etal-SfS1977}, biosensing
\cite{Liedberg-etal-S&A1983}, biological imaging
\cite{Okamoto-etal-PSP1990} and emerging new areas
\cite{Leung-etal-SAB1994}.

Enhanced electromagnetic fields boost nonlinear phenomena around the surface, which
are normally too tiny in magnitudes to be investigated. However, it
is difficult to see the effect of the surface response in
third-order nonlinear processes due to the strong bulk responses. This
can be remedied by the excitation of surface plasmons in a system
using coupling prisms in the Otto or Kretschmann configuration,
where surface plasmon polaritons are excited through attenuated total internal reflection
\cite{Yeatman-B&B1996}. The prism coupling scheme
provides a simple way to enhance the momentum of the incident light,
which is required for the excitation of surface plasmons
\cite{Torma-etal-RPP2015}.
A similar enhancement of nonlinear responses can be achieved by exciting
other wave modes confined near the surface such as slab waveguide modes.

Surface plasmons are usually generated near the frequencies of
visible light. As we approach the infrared region, the loss grows
rapidly. Recently, it has been shown that active photonic materials are promising
in photonics at infrared, ultraviolet and optical frequencies. The
gain provided by the emission of a gain medium sustains the
electromagnetic wave suffering high attenuation in the metal.
It has been demonstrated that it is possible to obtain net gain over
losses in a dielectric-metal-dielectric plasmonic waveguide, through
an optically pumped layer of fluorescent conjugated polymer adjacent
to the metal surface \cite{Gather-etal-NPh2010}.

In addition, gain has directly been measured in a gain medium
composed of optically pumped dye molecules, using the long-range
surface plasmon-polariton supported by a symmetric metal strip
waveguide \cite{De_Leon-etal-NPh2010}. Furthermore, recent
experiments have observed optical loss compensation effects in the
medium of randomly dispersed nanoshell particles. Coumarin C500 and
Rhodamine 6G (R6G) fluorescent dyes were encapsulated into the
dielectric shell \cite{Strangi-etal-PRL2011,De_Luca-etal-ACSN2011}.
It has been shown that metamaterials combined with a gain medium can
enhance effective gain due to the strong local field enhancement
incurred by metamaterials
\cite{Bergman-etal-PRL2003,Stockman-NPh2008}. There has been a study
of metamaterials capable of compensating loss, which are made of a
three-dimensional periodic array of nanospheres filled with R6G
inside the core \cite{Campione-etal-OME2011}. The utility of gain
medium has been expanded by employing Fano-resonant coupling between
a plasmonic shell and dielectric core, to cloaking and resonant
scattering \cite{Argyropoulos-etal-PRL2012}, third harmonic
generation \cite{Yang-etal-NL2015}, and optical pulling and pushing
\cite{Bian-etal-OE2017,Gao-etal-PRA2017}.

In this paper, we study the interplay between the linear and nonlinear gain effects and
the excitation of surface plasmons or a waveguide mode confined at the metal-dielectric boundary or inside the dielectric layer in the Otto configuration. The dielectric material
has Kerr-type optical nonlinearity, with a negative imaginary part
representing nonlinear gain. We also consider the influence of the negative imaginary
part of the linear dielectric permittivity.

Our theoretical method is based on the invariant imbedding method, using which we transform
the wave equation into a set of invariant imbedding equations
to create an initial value problem. This method
is a powerful tool to solve various kinds of singular and
nonlinear wave problems
\cite{KKim-etal-JKPS2001,KKim-etal-EPL2005,KKim-etal-OE2008,Phung-etal-JKPS2008}.
We introduce our formalism briefly in the following section.

\section{Model and Numerical Method}

For a $p$-polarized wave propagating in a layered structure,
the complex amplitude of the magnetic field, $H = H(z)$, satisfies
the wave equation
\begin{equation}\label{p-wave_eq}
{d^2 H \over dz^2} - {1 \over \epsilon(z)} {d \epsilon \over dz} {dH
\over dz} + \left[ k_0^2 \epsilon(z) - q^2 \right] H = 0,
\end{equation}
where $k_0$ ($ = \omega / c$) is the vacuum wave number and $\epsilon$ is the dielectric permittivity. When the wave is
incident from the region where $z>L$ and $\epsilon=\epsilon_1$ at an angle $\theta$ and propagates in the $xz$ plane, $q$ ($ = \sqrt{\epsilon_1
} k_0 \sin \theta$) and $p$ ($ =
\sqrt{\epsilon_1} k_0 \cos \theta$) are the $x$ and negative $z$ components of the wave vector.
We assume the layered structure lies in $0 \le z \le L$ and has a Kerr-type
nonlinearity in the dielectric permittivity. Then we have
\begin{equation}
\epsilon(z)=\begin{cases}
              \epsilon_1 & \text{if $ z > L $} \\
              \epsilon_L(z) + \alpha(z) \vert {\bf E}(z) \vert^2 & \text{if $ 0 \le z \le L $} \\
              \epsilon_2 & \text{if $ z < 0 $}
\end{cases},
\end{equation}
where $\epsilon_L (z)$ and $\alpha (z)$ are arbitrary complex
functions of $z$. We define the wave function in the incident region
as
\begin{equation}
H(z) = v \sqrt{\epsilon_1} \left[ e^{ip(L-z)+iqx} + r(L)
e^{ip(z-L)+iqx} \right],
\end{equation}
where $\vert v\vert^2$ ($\equiv w$) is the intensity of the electric field in the
incident wave and $r$ is the reflection coefficient.

We can transform the wave equation into a set of differential
equations for the reflection and transmission coefficients, $r$ and $t$,
and the imbedding parameter $w$ using the invariant imbedding method:
\begin{eqnarray}
{1 \over p} {dr(l) \over dl} & = & 2i {\epsilon(l) \over \epsilon_1}
r(l) - {i \over 2} \left[ {\epsilon(l) \over \epsilon_1} - 1 \right]
\left[ 1 - {\epsilon_1 \over \epsilon(l)}
\tan^2 \theta \right] \left[ 1 + r(l) \right]^2, \nonumber \\
{1 \over p} {dt(l) \over dl} & = & i {\epsilon(l) \over \epsilon_1}
t(l) - {i \over 2} \left[ {\epsilon(l) \over \epsilon_1} - 1 \right]
\left[ 1 - {\epsilon_1 \over \epsilon(l)}
\tan^2 \theta \right] \left[ 1 + r(l) \right] t(l), \nonumber \\
{1 \over p} {dw(l) \over dl} & = & {\rm Im} \left\{ 2 {\epsilon(l)
\over \epsilon_1} - \left[ {\epsilon(l) \over \epsilon_1} - 1
\right] \left[ 1 - {\epsilon_1 \over \epsilon(l)} \tan^2 \theta
\right] \left[ 1 + r(l) \right] \right\} w(l).
\end{eqnarray}
The dielectric permittivity is determined from the cubic equation
\begin{equation}
\epsilon(l) = \epsilon_L(l) + \alpha(l) w(l) \left[ {\epsilon_1^2
\over \vert \epsilon(l) \vert^2} \vert 1 + r(l) \vert^2 \sin^2 \theta + \vert 1 - r(l)
\vert ^2 \cos^2 \theta \right].
\end{equation}
Therefore, the invariant imbedding method solves an initial value
problem for a system of nonlinear differential equations and
the cubic polynomial is solved at each step to evaluate
$\epsilon(l)$ inside the structue.

The initial conditions for $r$, $t$ and $w$ are
\begin{eqnarray}
r(0) & = & {\epsilon_2 \sqrt{\epsilon_1} \cos \theta - \epsilon_1
\sqrt{\epsilon_2 - \epsilon_1 \sin^2 \theta} \over \epsilon_2
\sqrt{\epsilon_1} \cos \theta + \epsilon_1 \sqrt{\epsilon_2 -
\epsilon_1 \sin^2 \theta}}, \nonumber \\
t(0) & = & {2 \epsilon_2 \sqrt{\epsilon_1} \cos \theta \over
\epsilon_2 \sqrt{\epsilon_1} \cos \theta + \epsilon_1
\sqrt{\epsilon_2 -
\epsilon_1 \sin^2 \theta}}, \nonumber \\
w(0) & = & w_0.
\end{eqnarray}
The constant $w_0$ is chosen such that the final solution for $w(L)$
is the same as the physical input intensity. The reflectance $R$ is
given by $R = \vert r(L) \vert^2$ and the transmittance $T$ is
\begin{equation}
T = {\epsilon_1 \sqrt{\epsilon_2 - \epsilon_1 \sin^2 \theta}
\over \epsilon_2 \sqrt{\epsilon_1} \cos \theta} \vert t(L)\vert^2.
\end{equation}
If $\epsilon$ is real, there is no loss or gain and the
quantities $R$ and $T$ satisfy the law of conservation of energy, $R
+ T = 1$.

We can also calculate the field distribution inside the medium using the
invariant imbedding equation for the normalized magnetic field
amplitude $u(z) = H(z) / v$ is given as
\begin{equation}
{1 \over p} {\partial u(z,l) \over \partial l} = i {\epsilon(l)
\over \epsilon_1} u(z,l) - {i \over 2} \left[ {\epsilon(l) \over
\epsilon_1} - 1 \right] \left[ 1 - {\epsilon_1 \over \epsilon(l)}
\tan^2 \theta \right] [1 + r(l)] u(z,l),
\end{equation}
For a given $z$ ($0 < z < L$), $u(z,L)$ is obtained by integrating
this equation, together with the equations for $r(l)$ and
and $w(l)$, from $l = z$ to $l = L$ using the initial condition
$u(z,z) = 1 + r(z)$.

\section{Results and Discussion}

\begin{figure}[htbp]
\centering
\includegraphics[width=12cm]{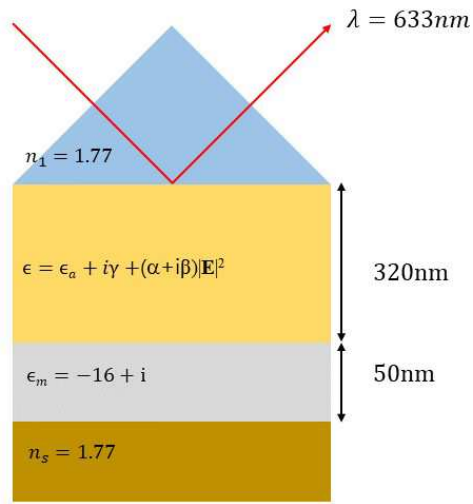}
\caption{\label{fig1} Schematic of the Otto configuration. }
\end{figure}

We are mainly interested in the interplay between the surface confined wave modes and nonlinear gain in the
Otto configuration (see Fig.~1). A $p$ wave is incident from a prism onto a nonlinear dielectric-metal bilayer, which lies on a dielectric substrate.
We assume that both the prism and the substrate have refractive indices of 1.77 and
the dielectric permittivity of the
nonlinear dielectric layer is given by
\begin{equation}
\epsilon =
\epsilon_a+i\gamma + \left(\alpha + i \beta\right) \vert {\bf E}\vert^2.
\end{equation}
We note that both the linear part of the permittivity and the third-order nonlinear optical susceptibility
are complex quantities.
We choose $\epsilon_a=1.46^2=2.1316$. The wavelength of the incident wave is 633 nm and
the dielectric permittivity of the metal layer is $-16+i$ corresponding to silver at that wavelength.
The thicknesses of the nonlinear dielectric and metal layers are 320 nm and 50 nm. Positive (negative) values of $\gamma$ and $\beta$ correspond to linear and nonlinear losses (gains) respectively. The critical incident angle $\theta_c$
in the linear case is equal to $55.57^\circ$.

\begin{figure}[htbp]
\centering
\includegraphics[width=8cm]{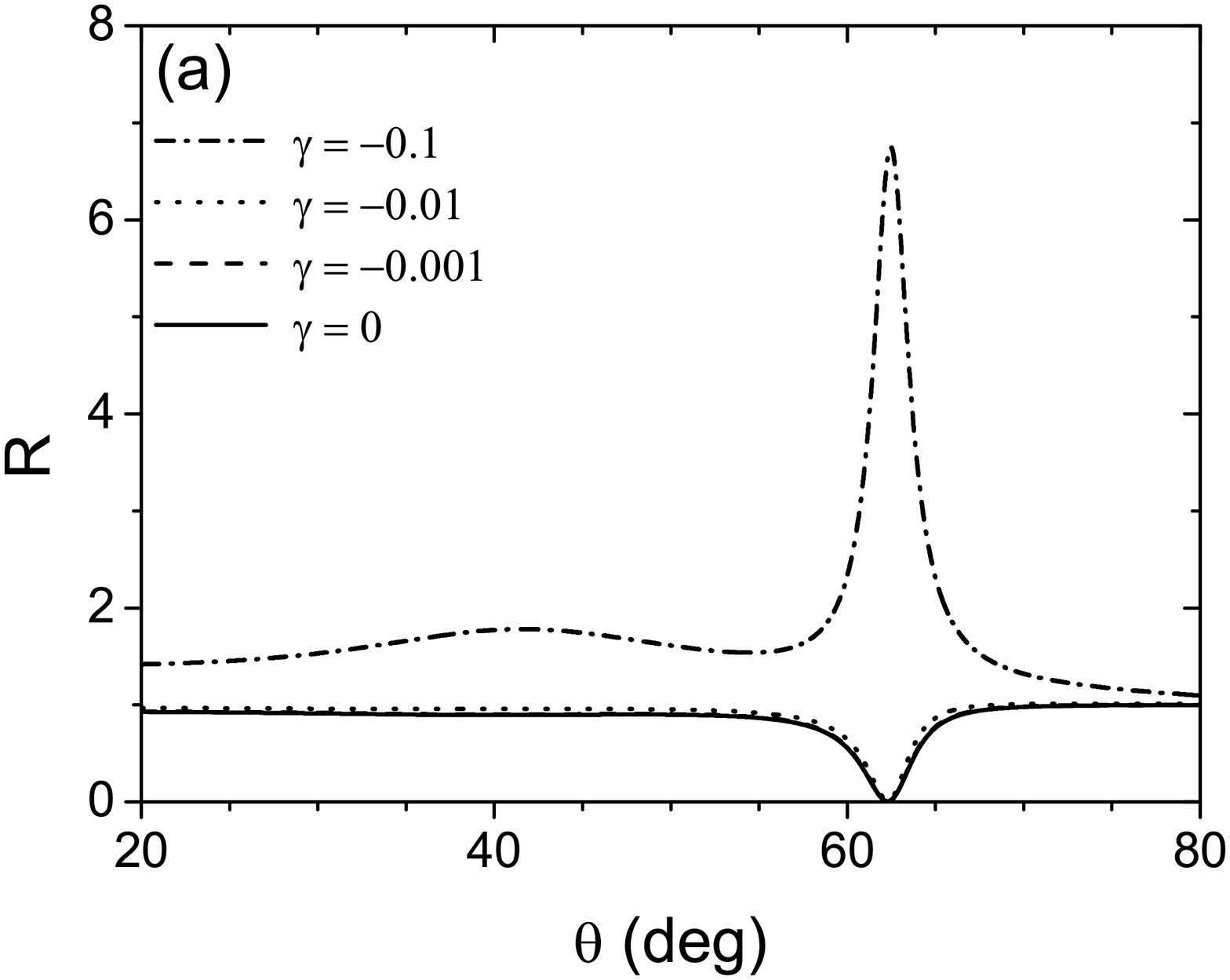}
\includegraphics[width=8cm]{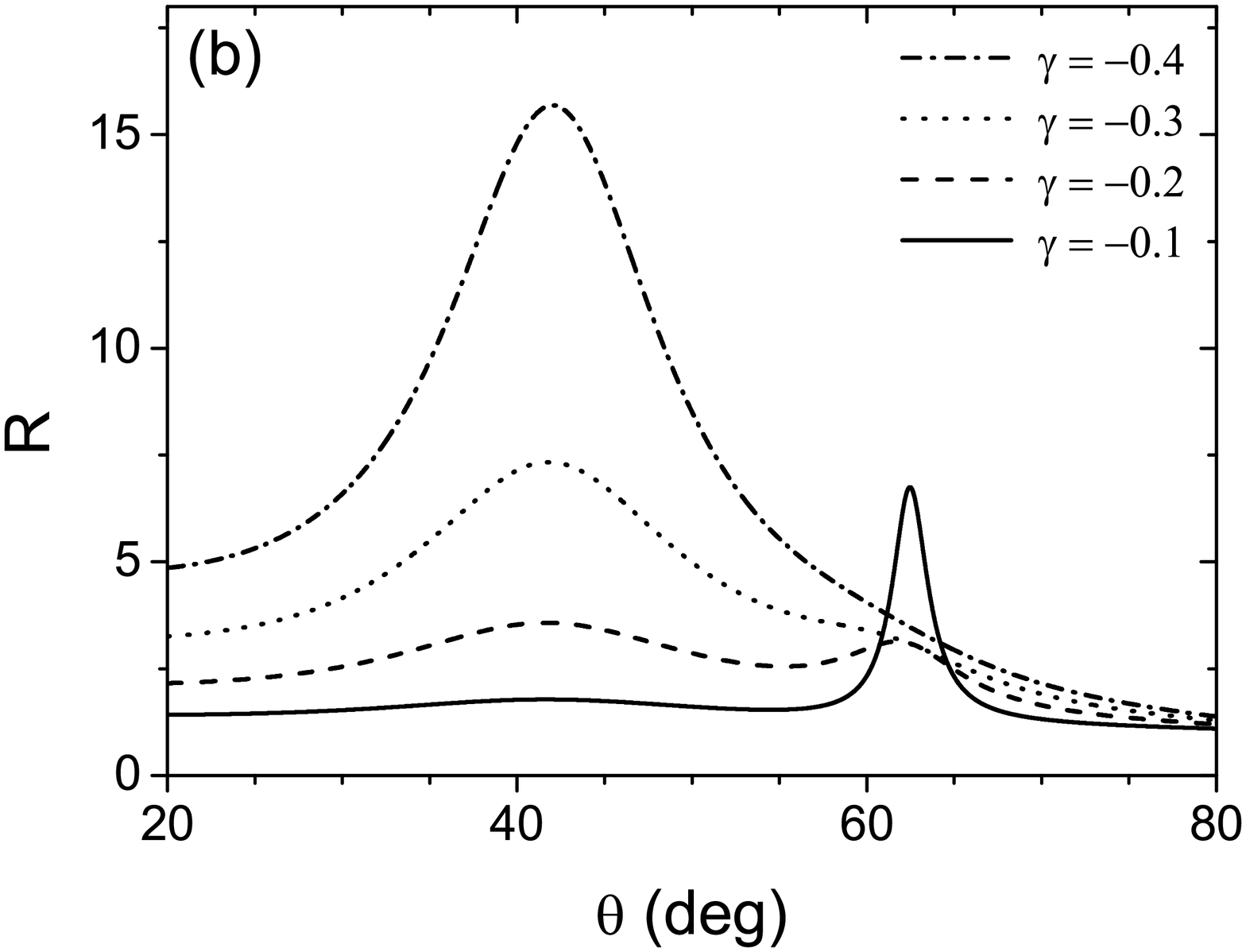}
\caption{\label{fig1} Reflectance vs. incident angle for various
values of the linear gain coefficient $\gamma$ when $\alpha=\beta=0$. (a) $\gamma = 0,~ -0.001,~ -0.01,~ -0.1$, (b) $\gamma = -0.1,~ -0.2,~
-0.3,~ -0.4$.}
\end{figure}

We first consider the situation where there is only linear gain.
In Fig.~2, we plot the reflectance $R$ versus incident angle
$\theta$ for different values of the linear gain coefficient
$\gamma$ when $\alpha=\beta=0$. In Fig.~2(a), we find that the influence of linear
gain is small when $\vert\gamma\vert \le 0.01$, since the reflectance curves
do not show much appreciable change. There is a dip in $R$
due to the excitation of surface plasmons at $\theta = 62.285^\circ$
for $\gamma = 0$ and at $\theta = 62.29^\circ$ for $\gamma =
-0.0001$ and $-0.01$. When $\gamma = -0.1$, the reflectance shows a peak
rather than a dip at $\theta =
62.47^\circ$ with the peak value ($R \approx 6.75$) much larger than 1. 
Thus the reflected power is larger than the incident power.
This over-reflection signifies
the amplification of the wave energy due to the interplay between the excitation of surface plasmons and the linear gain effect.
We also find that a new broad peak with the peak value of $R \approx 1.781$ occurs around $\theta = 41.61^\circ$. This peak is
due to the interplay between a waveguide mode inside the dielectric layer and linear gain.

In Fig.~2(b), we show the results for larger values of the gain
coefficient $\vert\gamma\vert$. We find that as $\vert\gamma\vert$
increases, the peak associated with surface plasmon excitation
(around $\theta = 62^\circ$) gradually decreases away, while the
peak associated with a waveguide mode (around $\theta = 42^\circ$)
gets strongly enhanced. We can see clearly that there are two peaks
when $\gamma = -0.2$. When $\gamma = -0.4$, $R$ reaches $15.6867$ at
$\theta = 42.025^\circ$. Since this angle is smaller than
$\theta_c$, refracted waves can propagate inside the dielectric
layer.

\begin{figure}[htbp]
\centering
\includegraphics[width=8cm]{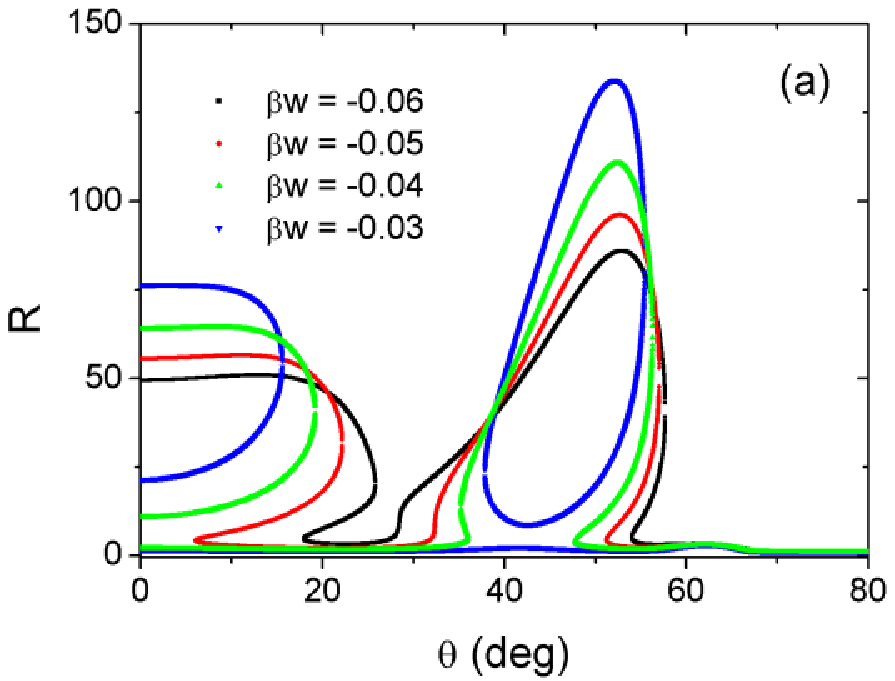}
\includegraphics[width=8cm]{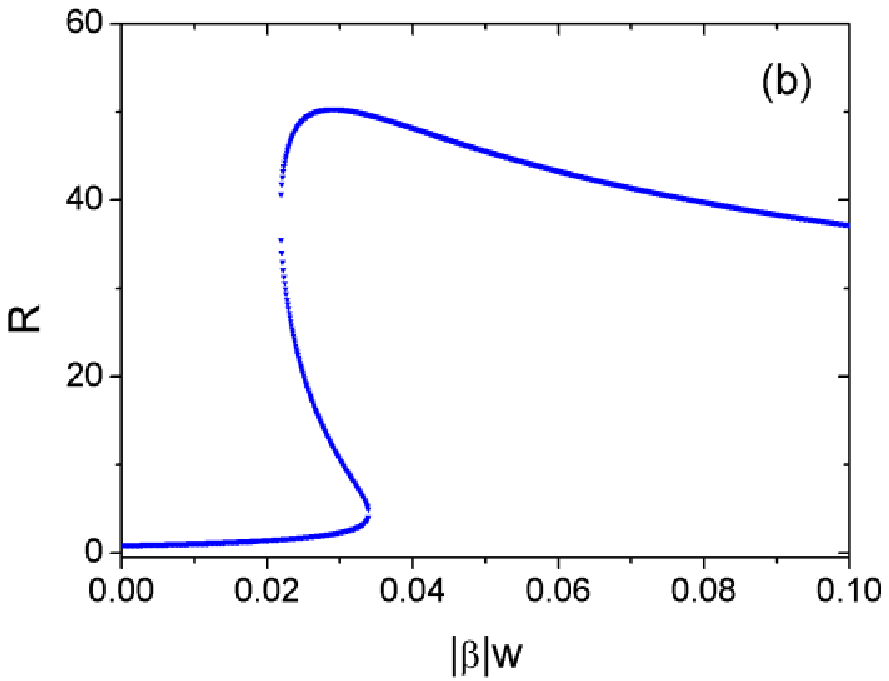}
\caption{\label{fig2} (a) Reflectance vs. incident
angle for various values of the nonlinear gain coefficient $\beta w$ and
(b) reflectance vs. nonlinear gain coefficient $\vert\beta\vert w$ for the
incident angle $\theta =40^\circ$, when $\gamma = 0$ and $\alpha =
0$.}
\end{figure}

Next, we consider the influence of nonlinear gain. A proper
dimensionless parameter measuring the strength of the nonlinear gain
effect is $\beta w$, where $w$ is the intensity of the incident
wave. For now, we turn off $\gamma$ and $\alpha$ and focus on the
effect of $\beta w$. In Fig.~3(a), we plot the reflectance versus
incident angle for various values of the nonlinear gain parameter
$\beta w$. As $\vert\beta\vert w$ is reduced from 0.06, the solution
curve displaying multistability goes through successive deformation.
For $\beta w = -0.03$, the curve has three separate parts, one of
which is a closed curve located between $\theta\approx 38^\circ$ and
$\theta\approx 55^\circ$. We can distinguish the stable solutions from the unstable ones by
plotting the reflectance versus $\beta w$. In Fig.~3(b),
we show the reflectance curve for $\theta =
40^\circ$ when the nonlinear gain parameter $\vert\beta\vert w$  is varied. 
We notice that there is a region where there exist three solutions for a given value of $\vert\beta\vert w$. 
It is
well-known that nonlinear resonance problems have multiple solutions
with alternating stable and unstable branches for a given intensity
of incident light \cite{Ovchinnikov-etal-JETP2001}. The unstable
branch is located between the stable branches at its bottom and top
positions. This is a usual mechanism for hysteresis. The change in
the number of multistable solutions for varying $\theta$ is not
monotonic.

\begin{figure}[htbp]
\centering
\includegraphics[width=9cm]{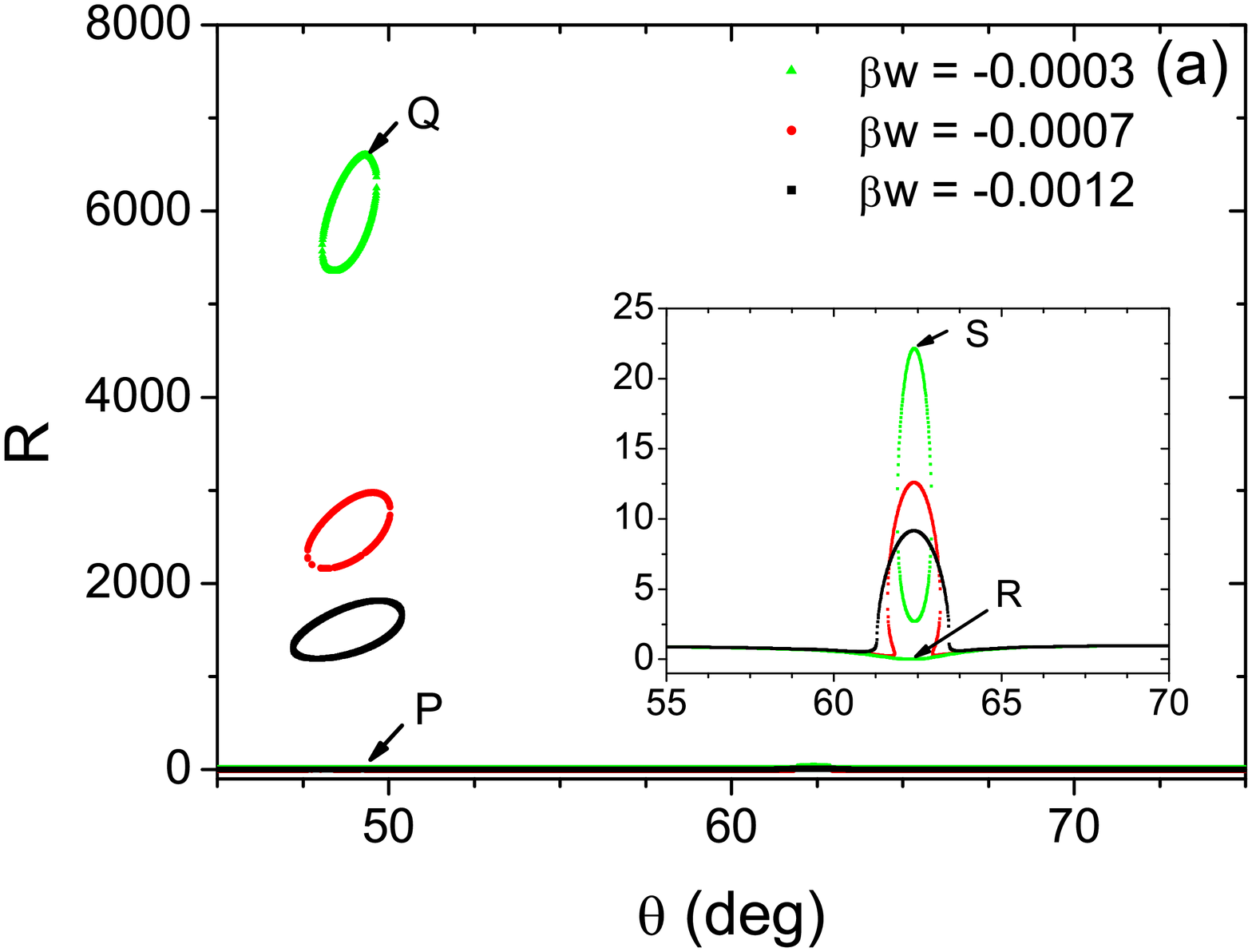}
\includegraphics[width=9cm]{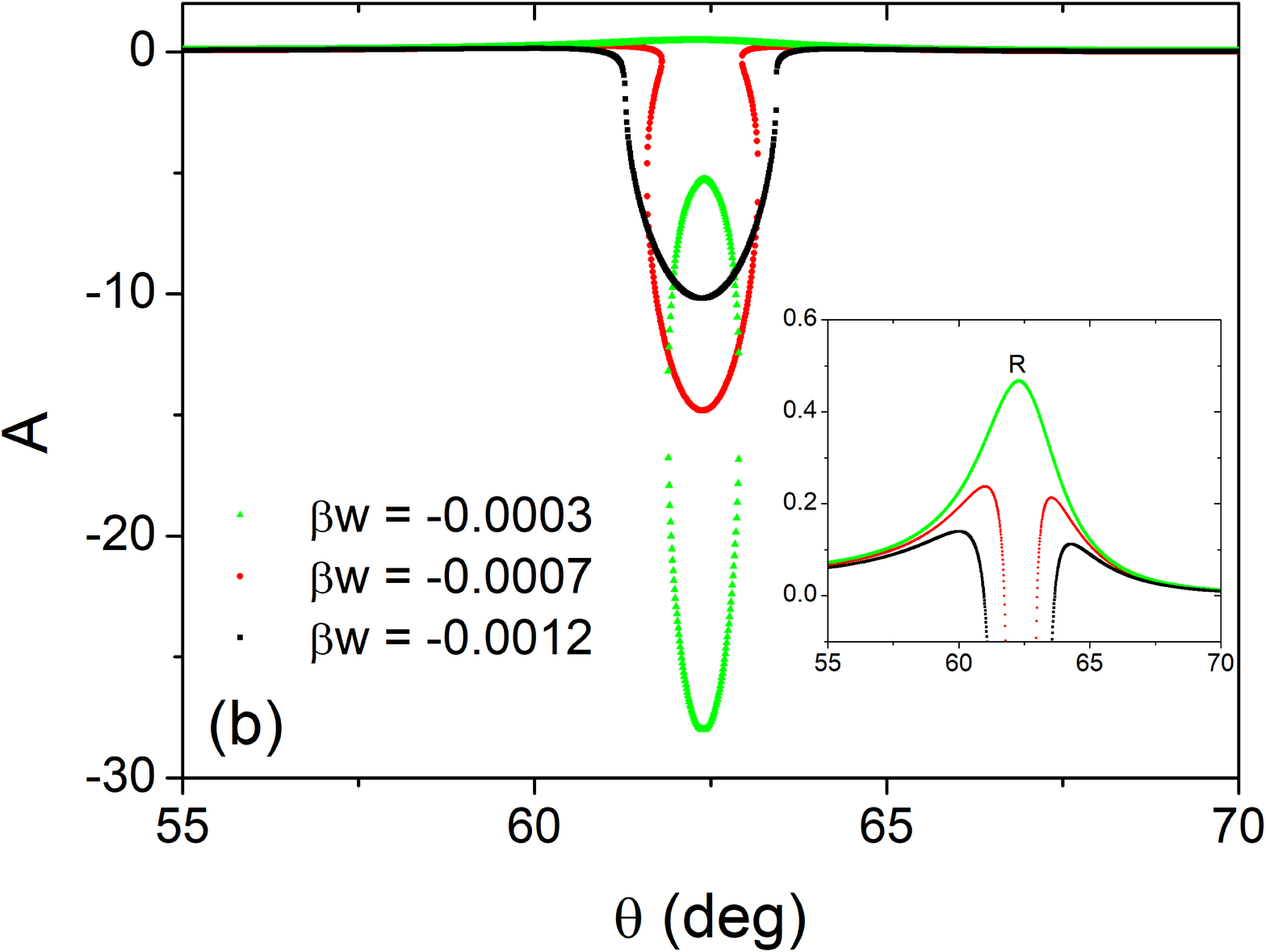}
\caption{\label{fig3}
(a) Reflectance $R$ and (b) absorptance $A$ plotted vs. incident
angle for various values of the nonlinear
gain coefficient $\beta w$, when $\gamma = 0$ and $\alpha = 0$.
Insets: the expanded view of some parts of each plot. }
\end{figure}

In Fig.~4(a), the reflectance curves for smaller values of
$\vert\beta\vert w$ are shown. The values of $R$ become very large,
reaching about 6601 for $\beta w = -0.0003$. In the range of the
incident angle where such huge amplification takes place, a
propagating wave travels inside the dielectric layer. Our explanation
about the possible physical mechanism will be given in the
discussion of Fig.~5 later.

In the inset of Fig.~4(a), the curves around $\theta = 62.28^\circ$
are expanded to show their behavior in detail. In this range of
the incident angle, an evanescent wave is amplified
through the excitation of surface plasmons. When $\beta w = -0.0007$,
the reflectance takes a local maximum around $\theta = 62.28^\circ$.
Since the curve is deformed to form a partial loop (in the shape of
$\Omega$) around this angle, there are two optical bistability
regions, which occur in the small ranges of incident angles in $61.6^\circ
- 61.81^\circ$ and $62.95^\circ - 63.17^\circ$. When the incident angle
goes through these regions of optical bistability, the reflectivity
jumps up from a small to a larger value, or vice versa. When the
nonlinear gain parameter is about $-0.0012$, the optical bistability
cannot be observed and the maximum of the reflectance is smaller
than the value in the previous case where the nonlinear gain
parameter is about $-0.0007$.

In Fig.~4(b), we plot the absorptance $A$ ($= 1 - R - T$). In most of
our calculations, $A$ is negative due to the
fact that $R$ and $T$ are greater than 1. In
the inset of Fig.~4(b), the plot is expanded around $A = 0$. At the
point R corresponding to $\theta = 62.29^\circ$, $A \approx 0.467>0$ when $\beta w = -0.0003$. 
This case corresponds to the point R in
Fig.~4(a), where $R$ and $T$ are less than 1. When $\vert\beta\vert w \ll 1$,
the electromagnetic fields associated with the surface plasmon are weak and confined
close to the nonlinear dielectric-metal interface. Because of this, the damping effect due to the metal layer
can be bigger than the amplification effect due to the dielectric layer and  
we have $0 < A < 1$. As $\vert\beta\vert w$ increases, the surface
plasmon fields have bigger amplitude and the amplification effect dominates metallic damping, resulting in $A < 0$ [see Fig.~5(b)]. 
Therefore, as
the incident intensity $w$ is increased up and decreased down [see
Fig.~3(b)], the system makes a transition between damping and amplification. This may lead to a new device
which allows a system to have a self-control to
prevent thermal damage.

\begin{figure}[htbp]
\centering
\includegraphics[width=8cm]{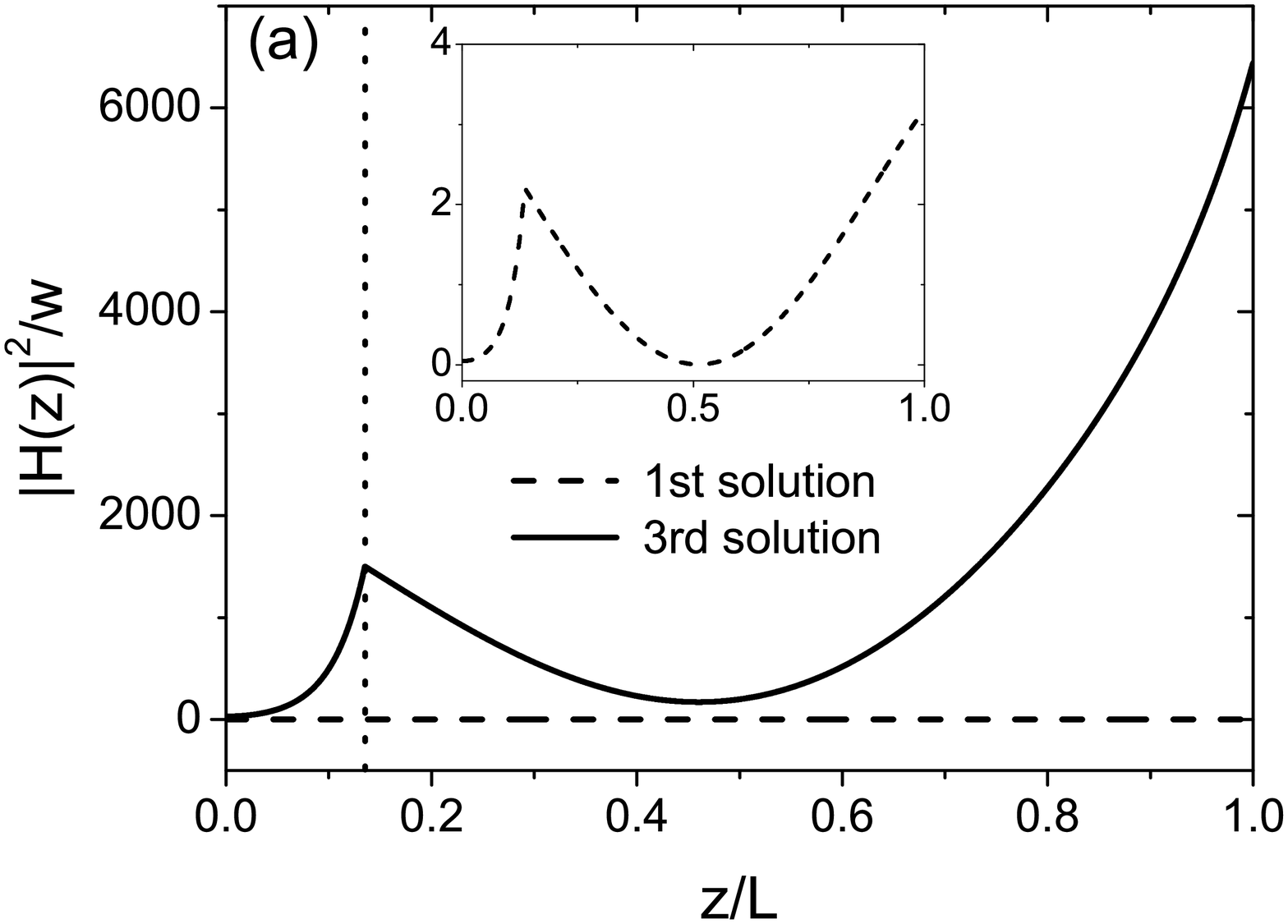}
\includegraphics[width=8cm]{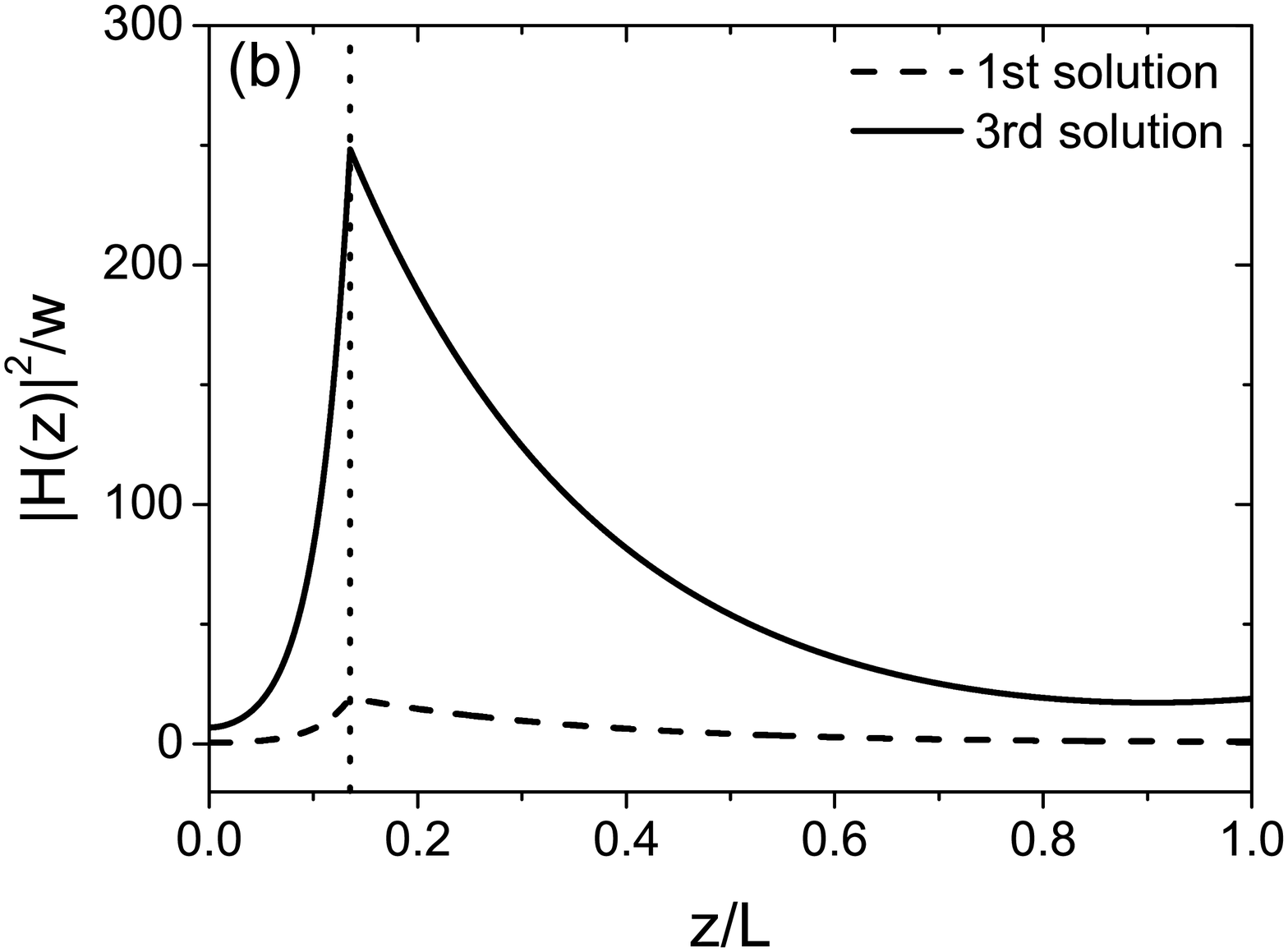}
\caption{\label{fig4} 
Normalized field profiles inside the metal/dielectric bilayer of thickness $L$ ($=370$ nm) at the
points P, Q, R and S indicated in Fig.~4(a). (a) 1st solution (point P) and 3rd
solution (point Q) at $\theta = 49.32^\circ$. Inset: the expanded view of the 1st solution. (b) 1st solution
(point R) and 3rd solution (point S) at $\theta = 62.29^\circ$. The vertical dotted line indicates
the position of the metal/dielectric boundary. The wave is incident from $z>L$ onto the dielectric layer side.}
\end{figure}

In Fig.~5, we show the magnetic field profiles corresponding to the
points P, Q, R and S indicated in Fig.~4(a). In Fig.~5(a), the first
solution (P) and the third solution (Q) at $\theta = 49.32^\circ$
are shown. In the inset of Fig.~5(a), an expanded view of
the first solution is shown. The third solution is amplified strongly near
the input side. As we have seen in Fig.~2, this resonance occurs at
incident angles less than the critical angle in a medium with strong
gain. Nonlinear gain is not homogeneous in the medium due to the
inhomogeneous electric field. When a light with high intensity is
incident on the system, the refractive index near the input side
will increase, thus the propagating waves are reflected before
reaching the other end. Hence in order to obtain strong enhancement,
the nonlinear gain must not be too high.

The situation in Fig.~5(a) shows that some sort of Fabry-Perot
cavity is formed in the dielectric layer, where propagating waves
travel between the input face and the metal-dielectric interface. This will lead to
strong amplification. As the local field is enhanced, the nonlinear
gain $\beta w$ gets larger, resulting in the growth of resonance.
Such enormous enhancement looks promising, but the high electric
field boosted by three orders of magnitude may cause damage in the
thin dielectric or metal film due to Joule heating
\cite{Sambles-etal-JMO1988} even in the Otto configuration.
In Fig.~5(b), the first solution corresponds to the point R, which is
the point where the damping ($A > 0$) takes the maximum value
shown in the inset of Fig.~4(b) at $\theta = 62.29^\circ$.

\begin{figure}[htbp]
\centering
\includegraphics[width=9cm]{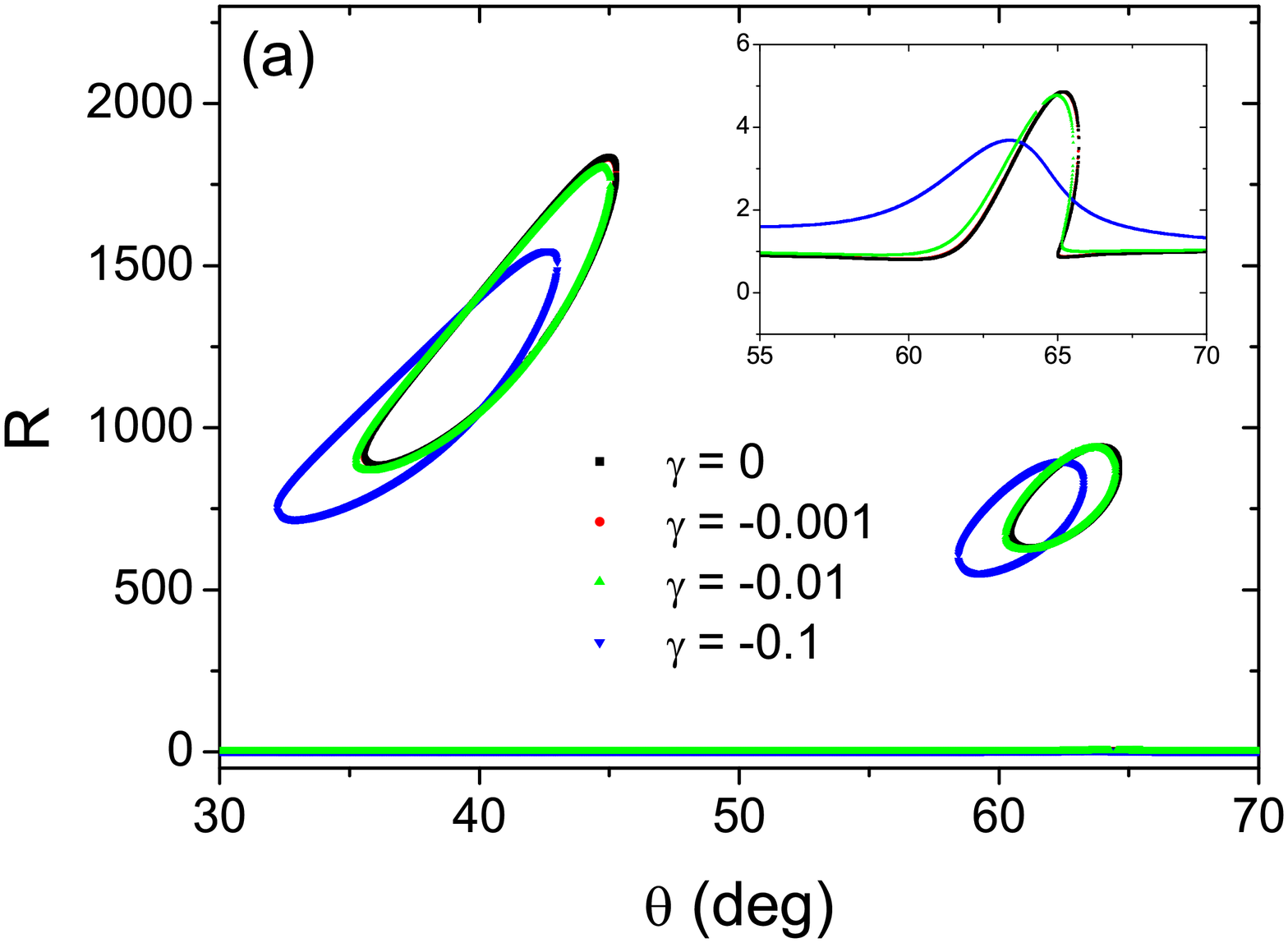}
\includegraphics[width=9cm]{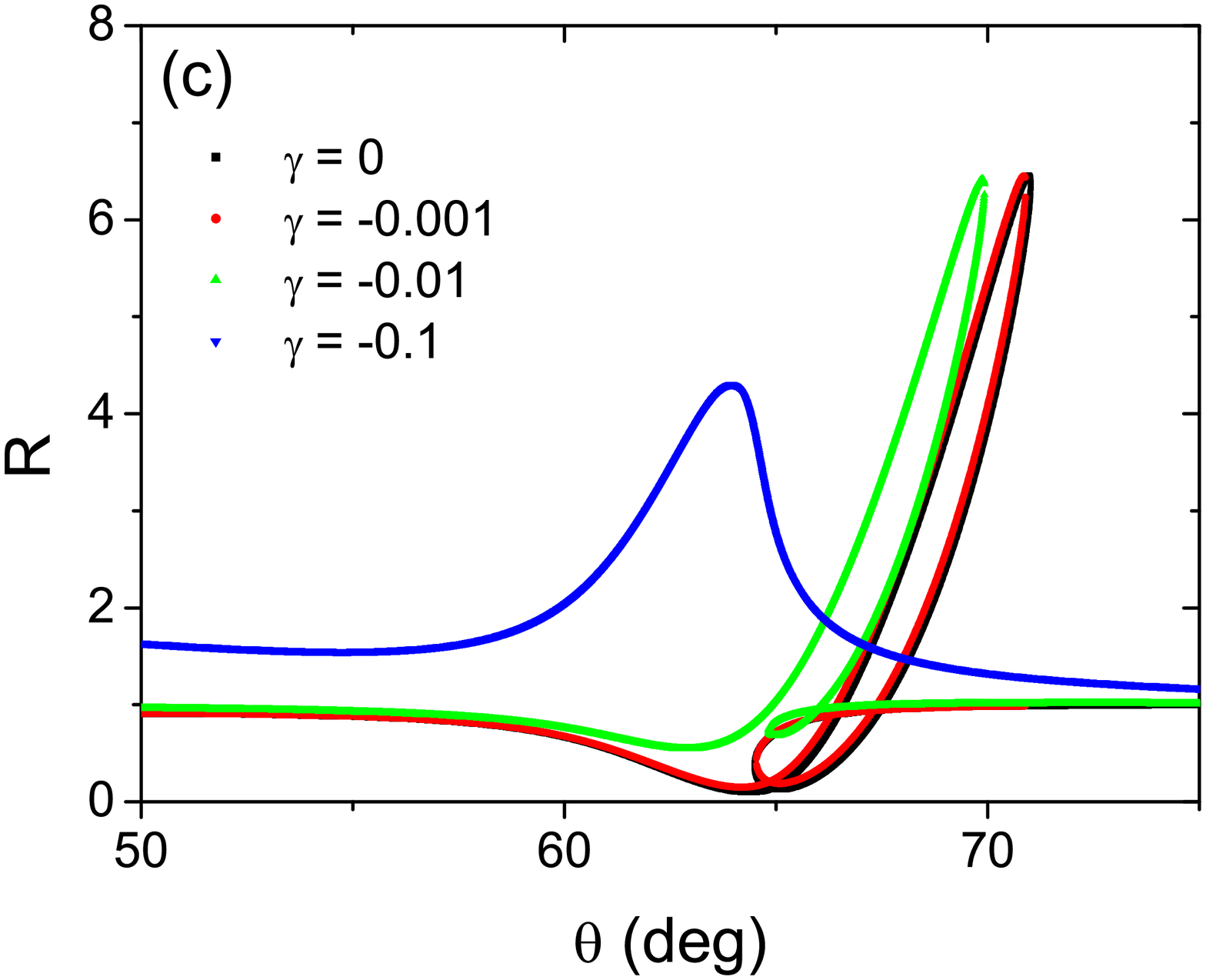}
\caption{\label{fig5} 
Reflectance vs. incident angle for various
values of the linear gain coefficient $\gamma$, when (a) $\alpha w =
0.0028$ and $\beta w = -0.0028$, (b) $\alpha w = 0.0028$ and $\beta
w = -0.0007$. Inset: the expanded view of some part of (a).}
\end{figure}

So far we have considered the effects of the gain parameters
$\gamma$ and $\beta w$ separately. Now we consider the case where
the dielectric layer has both focusing nonlinearity and nonlinear
gain. In Fig.~6(a), we plot the reflectance versus incident angle
for several values of $\gamma$, when $\alpha w = 0.0028$ and $\beta
w = -0.0028$. For all values of $\gamma$, there are two closed
solution curves with very large $R$ around $\theta = 40^\circ$ and
$\theta = 62^\circ$. In addition, there exist solutions with much
smaller values of $R$ in the interval $55^\circ<\theta<70^\circ$,
which are shown in the inset. The optical bistability can be
observed in this case when $\vert\gamma\vert<0.1$. When $\gamma =
-0.1$, no optical bistability is observed. It appears that the
self-focusing nonlinearity $\alpha w$ quenches the amplification
effect due to the linear and nonlinear gain, while multistability
occurs in a wider range of the incident angle. It is well-known
that Kerr-type nonlinearity can induce severely multistable solutions \cite{KKim-etal-OE2008}. 
The reflectance and the transmittance change substantially
for a very small change of the nonlinearity parameter $\alpha w$ and
the field profiles inside the system are highly nonuniform.
When the focusing nonlinearity is turned on in Fig.~6, the
propagating waves in the dielectric layer encounter inhomogeneous fields
varying in a complicated manner, even for a small change in the input
intensity. Thus the Fabry-Perot-type amplification as in Fig.~5(a)
does not occur, as is evident in Fig.~6(b). Such complex
landscape of fields would also degrade the propagation of surface
plasmons. If we reduce the strength of nonlinear gain such that
$\alpha w = 0.0028$ and $\beta w = -0.0007$, the influence of
$\alpha w$ becomes stronger and amplification is greatly reduced. We
observe that complicated patterns of multistability appear in this
case. When $\gamma = -0.1$, no multistability is observed.

In a saturable nonlinear gain medium, saturation occurs when the amplification
of surface plasmon polaritons is compensated by the nonlinear losses
\cite{Marini-etal-OL2009}. In this case, taking the saturation effect into
account, the dielectric constant may be written as 
\begin{eqnarray}
\epsilon =
\epsilon_L + {\alpha_s \vert {\bf E}\vert^2 \over 1 + \beta_s \vert {\bf E}\vert^2},
\end{eqnarray} 
where
$\epsilon_L$ is the linear part and $\alpha_s$ and $\beta_s$ are the
coefficients for the nonlinear term. For $\beta_s \ll 1$, it can be
approximated to 
\begin{equation}
\epsilon \approx \epsilon_L + \alpha_s \vert {\bf E}\vert^2 -
\alpha_s \beta_s \vert {\bf E}\vert^4.
\end{equation}
Thus a higher order term with the opposite
sign to the Kerr-nonlinear term appears, introducing loss. Under
this condition, the field enhancement on the boundaries of the dielectric
layer should be limited. On the other hand, our model does not have
loss built inside since it is in the form $\epsilon = \epsilon_L +
\alpha \vert {\bf E}\vert^2$. However, the Fabry-Perot-type amplification in
Fig.~5(a) has its own limit, because the field in the middle of
the dielectric layer cannot grow faster than the field at the boundaries.
Thus as $\vert\beta\vert w$ grows, the amplification is reduced.

In order to observe the effects discussed in this paper,
we have found that we need to have the nonlinear gain
parameter $\vert\beta\vert w$ in the order of magnitude of $10^{-4}$.
We point out that systems with large nonlinear gain parameters have been studied
in the area of soliton transmission in fibers \cite{a}.
Though we do not currently have a realistic model relating the nonlinear gain parameter with
material dielectric functions, we speculate that it may be possible to fabricate systems with large nonlinear gain by
combining materials with large Kerr coefficients and gain media, using the scheme of metamaterials research.

\section{Conclusion}

In this paper, we have studied the interplay between the surface confined wave modes such as surface plasmons and waveguide modes
and the linear and nonlinear gain effects in the Otto configuration. In the linear case, we have
found that a large reflectance peak is generated due to this interplay and the incident angle at which the peak occurs shifts from the angle associated with surface plasmons
to that associated with a waveguide mode, as the gain parameter increases.
When the nonlinear gain is nonzero, we have found that there appear strong multistability phenomena near the incident angles associated with
surface plasmons and a waveguide mode. We have shown that the reflectance curve can display topologically complicated multistability structure.
When the nonlinear gain parameter takes a small optimal value, we have found that a giant amplification of the reflectance by three orders of magnitude can occur near the incident angle associated with a waveguide mode.
We have also found that there exists
a range of the incident angle where the wave is dissipated in the presence of gain and suggested that this can provide
 a possible new technology for thermal
control in the subwavelength scale.

\section*{Funding}
National Research Foundation of Korea Grant funded by the Korean Government (NRF-2015R1A2A2A01003494).

\end{document}